\documentclass[twocolumn,aps,showpacs,prb,tightenlines,amsmath,amssymb]{revtex4}
\usepackage{graphicx}
\usepackage{amssymb}
\usepackage{dcolumn} 
\usepackage{amsmath}
\usepackage{bm}
\usepackage{colordvi}

\makeatletter

\newcommand{\Rmnum}[1]{\expandafter\@slowromancap\romannumeral #1@}
\makeatother

\begin{document}

\title{Dynamics of photoexcited carriers in graphene}
\author{B. Y. Sun} 
\author{Y. Zhou}
\author{M. W. Wu}
\thanks{Author to  whom correspondence should be addressed}
\email{mwwu@ustc.edu.cn.}
\affiliation{Hefei National Laboratory for Physical Sciences at
  Microscale and Department of Physics,
University of Science and Technology of China, Hefei,
  Anhui, 230026, China}
\date{\today}

\begin{abstract}
The nonequilibrium dynamics of carriers and phonons in graphene is
investigated by solving the microscopic kinetic equations with the carrier-phonon and carrier-carrier 
Coulomb scatterings explicitly included. The Fermi distribution of hot
carriers are found to be established within 100~fs and the temperatures of
electrons in the conduction and valence
  bands are very close to each other, even when the 
excitation density and the equilibrium density are comparable, thanks to the
strong inter-band Coulomb scattering. Moreover, the
temporal evolutions of the differential transmission obtained from our
calculations agree with the experiments by Wang {\it et al.} 
[Appl. Phys. Lett. {\bf 96}, 081917 (2010)] and Hale {\it et al.} [Phys. Rev. B
{\bf 83}, 121404 (2011)] very well, with two distinct differential transmission relaxations presented.
We show that the fast relaxation is due to the rapid carrier-phonon thermalization
and the slow one is mainly because of the slow decay of hot phonons.
In addition, it is found that the temperatures of the hot phonons in different branches are different
and the temperature of hot carriers can be even lower than that of the hottest
phonons. Finally, we show that the slow relaxation rate exhibits a mild
valley in the excitation density dependence and is linearly dependent on the probe-photon energy.
\end{abstract}

\maketitle

\section{INTRODUCTION}

The unique electrical and optical properties of graphene,
e.g., large mobility, long coherent length and exceptionally low electrical
noise, make it a promising material for the development of
nanoscale devices.\cite{Novoselov,Novoselov2,Castro,Schedin,FWang} The performance of many such   
devices depends critically on the dynamic properties of carriers and
phonons. Therefore, a thorough understanding of these properties is
essential.

The time-resolved optical pump-probe measurement is a powerful tool widely used to
probe the ultrafast dynamics of photoexcited carriers and has been applied
extensively to graphene lately.\cite{Dawlaty,Hale,Wang,Huang} 
In these works, a fast differential transmission (DT) relaxation of several hundred femtoseconds, followed by a
slower picosecond relaxation were observed.\cite{Dawlaty,Hale,Wang,Huang} Dawlaty {\it et al.}\cite{Dawlaty} 
suggested that the fast one is due to equilibrating of carriers 
through the carrier-carrier scattering and the slow one is related to the cooling
of the hot-carrier distribution through the carrier-phonon scattering. However, both
experimental\cite{Kampfrath,Breusing} and theoretical works\cite{Butscher} show
that the photoexcited carriers lose most energy to optical 
phonons within the time scale of the fast relaxation (about
500~fs) and the slow relaxation rate is in the same 
order of the hot-phonon decay rate obtained from the time-resolved Raman
spectroscopy.\cite{Kang,Yan} Therefore, the fast
relaxation is supposed to be associated with the rapid carrier-phonon
thermalization and the slow one to the relaxation of 
hot phonons through the phonon-phonon scattering.\cite{Hale,Wang,Huang} 
However, the theoretical investigations on this problem in the
literature are performed by using the coupled rate equations which calculate
the energy transferred among the carriers, phonons and environment.\cite{Wang,Hale}
This method is based on the ansatz that the carrier-carrier Coulomb scattering
is very strong so that the carrier Fermi distribution can be
established very rapidly. This should be
verified especially at high photo-excited carrier
density in which the screening is strong. Moreover, in their model, the
 temperatures of electrons in the conduction and valence 
 bands are assumed to be always identical after buildup of the Fermi
distribution. This is
reasonable when the excitation density is much larger than the
equilibrium carrier density, since the distributions of
electron and hole are almost
identical in this case. Nevertheless, it should be 
examined when these two  densities are comparable.
In addition, in the previous investigations, the contribution of
the remote-interfacial (RI) phonons is neglected and the carrier-phonon
scattering matrices for the longitudinal and 
transverse optical phonons near the $\Gamma$ point are set to be
identical.\cite{Ando,Piscanec,Lazzeri} The influence of the missing physics,
especially the different temperatures in various phonon 
branches from different carrier-phonon scattering strengths, is still
unclear. All these questions suggest that a detailed theoretical investigation
from a microscopic approach is essential.

In this paper, we investigate the nonequilibrium dynamics of carriers and phonons in 
graphene via the microscopic kinetic equation approach with
  the carrier-phonon and carrier-carrier Coulomb scatterings explicitly
included. The temporal evolutions of the carrier distribution  and the 
phonon number as well as the DT are obtained numerically. We find that the
hot-carrier Fermi distribution is established within less than 100~fs.
Furthermore, due to the strong inter-band Coulomb scattering, the temperatures
of electrons in conduction and valence bands are shown to be very close to each
other even when the excitation and equilibrium densities are comparable.
It is also shown that the calculated DTs have good agreement with 
  the experimental data by Hale {\it et al.}\cite{Hale} and Wang {\it et
    al.}\cite{Wang} for different graphene layer
  numbers and excitation densities.  Moreover, our
calculations provide strong evidence to the claim in the previous experimental
works\cite{Hale,Wang,Huang} that the fast relaxation of the
DT is due to the carrier-phonon thermalization and the slow one mainly comes from the hot-phonon decay.
Furthermore, it is shown that
due to the different carrier-phonon scattering strengths, the temperatures of hot phonons in 
different branches are different. Moreover, the temperature of carriers
can be even {\em lower} than that of the hottest phonon. Finally, it is discovered that the
slow relaxation rate exhibits a mild valley in the excitation-density dependence
and depends linearly on the probe-photon energy.

This paper is organized as follows. In Sec.~{\Rmnum 2}, we set up the model and
lay out the kinetic equations. In Sec.~{\Rmnum 3} the results obtained
numerically from the kinetic equations are presented. We summarize in
Sec.~{\Rmnum 4}.

\section{MODEL AND FORMALISM}
We start our investigation from  graphene on SiO$_2$ or SiC substrates. 
Exploiting the nonequilibrium Green's function approach, the kinetic equations of the
 carriers can be constructed as\cite{wuReview,yzhou,PZhang,PZhang2}
\begin{equation}
  \partial_t{f}_{\mu{\bf k}\nu}=\left.\partial_t{f}_{\mu{\bf
        k}\nu}\right|_{\rm ee}+\left.\partial_t{f}_{\mu{\bf k}\nu}\right|_{\rm
    ep}+\left.\partial_t{f}_{\mu{\bf k}\nu}\right|_{\rm ei}.
\label{KEE}
\end{equation}
Here $\mu=+(-)$ represents the K(K$^\prime$) valley, ${\bf k}$ stand for the
wave vectors relative to the K or K$^\prime$ points and $f_{\mu{\bf
    k}\nu}$  represent the electron distribution functions in the
conduction ($\nu=+$) or valence ($\nu=-$) bands. $\left.\partial_t{f}_{\mu{\bf k}\nu}\right|_{\rm
  ee}$ describe the carrier-carrier Coulomb scattering terms and
$\left.\partial_t{f}_{\mu{\bf k}\nu}\right|_{\rm ep}$ give the carrier-phonon scattering terms
including the scatterings between carriers and acoustic, optical as well as RI phonons. In this investigation, we assume
that the initial carrier distribution after the pumping is isotropic. Therefore,
the carrier-impurity scattering terms $\left.\partial_t{f}_{\mu{\bf
      k}\nu}\right|_{\rm ei}$ are always zero. 
The carrier-carrier scattering terms can be written as ($\hbar\equiv1$
throughout this paper) 
\begin{eqnarray} 
  \left.\partial_t{f}_{\mu{\bf k}\nu}\right|_{\rm ee}&=&-4 \pi  \sum_{{\bf
      k^\prime}\nu^\prime\mu^\prime}\sum_{{\bf k}_1\nu^\prime_1\nu^\prime_2} I_{{\bf
      k}\nu,{\bf k^\prime}\nu^\prime}I_{{\bf
    k_1+k-k^\prime}\nu^\prime_1,{\bf k_1}\nu^\prime_2}\nonumber \\
&&\hspace{-0.9cm}\mbox{}\times|V^{\nu\nu^\prime}_{{\bf
    k},{\bf k^\prime}}|^2\delta(\varepsilon_{{\bf
    k^\prime}\nu^\prime}-\varepsilon_{{\bf
    k}\nu}+\varepsilon_{{\bf k_1+k-k^\prime}\nu^\prime_1}-\varepsilon_{{\bf
    k_1}\nu^\prime_2}) \nonumber \\ 
&&\hspace{-0.9cm}\mbox{}\times\big(f^>_{\mu^\prime{\bf
    k_1+k-k^\prime}\nu^\prime_1}f^<_{\mu^\prime{\bf
    k_1}\nu^\prime_2}f^>_{\mu{\bf k^\prime}\nu^\prime}f^<_{\mu{\bf 
    k}\nu}-f^<_{\mu^\prime{\bf
    k_1+k-k^\prime}\nu^\prime_1}\nonumber\\
&&\hspace{-0.9cm}\mbox{}\times f^>_{\mu^\prime{\bf
    k_1}\nu^\prime_2}f^<_{\mu{\bf
    k^\prime}\nu^\prime}f^>_{\mu{\bf k}\nu}\big),
\end{eqnarray}
in which $I_{{\bf
    k}\nu,{\bf k^\prime}\nu^\prime}=\frac{1}{2}[1+\nu\nu^\prime\cos(\theta_{\bf
  k}-\theta_{\bf k^\prime})]$, $f^<_{\mu{\bf k}\nu}\equiv f_{\mu{\bf k}\nu}$,
$f^>_{\mu{\bf k}\nu}\equiv 1-f_{\mu{\bf k}\nu}$ and $\varepsilon_{{\bf k}\nu}= \nu
v_{\rm F}k$ with ${v_{\rm F}}$ being the Fermi velocity.
$V^{\nu\nu^\prime}_{{\bf k,k^\prime}}$ denotes the screened Coulomb potential under the random phase
approximation\cite{Haug,Hwang2,Hwang} 
\begin{equation}
V^{\nu\nu^\prime}_{{\bf k+q},{\bf k}}=V^0_{\bf
    q}/[{1-V^0_{\bf q}\Pi({\bf q},\varepsilon_{{\bf
    k+q}\nu}-\varepsilon_{{\bf k}\nu^\prime})}],
\end{equation}
in which $V_{\bf q}^0=2\pi v_Fr_s/q$ is
the two-dimensional bare Coulomb potential with $r_s$ being the dimensionless
Wigner-Seitz radius\cite{Adam,Hwang2,Adam2,Hwang,Fratini} and $\Pi({\bf q},\omega)$ is
given by\cite{Hwang,Wunsch,XFWang,Ramezanali}  
\begin{equation}
\Pi({\bf q},\omega)=\sum_{\mu \nu\nu^\prime {\bf k}}2I_{{\bf k}\nu,{\bf
    k+q}\nu^\prime}\frac{f_{\mu {\bf
    k}\nu}-f_{\mu {\bf
    k+q}\nu^\prime}}{\varepsilon_{{\bf
      k}\nu}-\varepsilon_{{\bf k+q}\nu^\prime}+\omega+i0^+}.
\end{equation}
The carrier-phonon scattering terms are given by
\begin{eqnarray}
\hspace{-0.1cm}\left.\partial_t{f}_{\mu{\bf k}\nu}\right|_{\rm
    ep}&=&-2 \pi\hspace{-0.1cm}\sum_{{\bf
      k^\prime}\mu^\prime\nu^\prime\atop\lambda,\pm}\hspace{-0.1cm}|M^{\lambda\mu\mu^\prime}_{\bf 
    k\nu,k^\prime\nu^\prime}|^2\delta(\varepsilon_{{\bf
      k^\prime}\nu^\prime}-\varepsilon_{{\bf
      k}\nu}\pm\omega_{{\bf k-k^\prime}\lambda})\nonumber\\
  &&\hspace{-1.7cm}\mbox{}\times\big(f^>_{\mu^\prime{\bf 
      k^\prime}\nu^\prime}f^<_{\mu{\bf k}\nu}n^\pm_{{\bf 
      {\bf k-k^\prime}}\lambda}-f^<_{\mu^\prime{\bf k^\prime}\nu^\prime} f^>_{\mu{\bf
      k}\nu}n^\mp_{{\bf k-k^\prime}\lambda}\big).
 \end{eqnarray}
Here $\lambda$ is the phonon branch index and $\omega_{{\bf q}\lambda}$ is
the corresponding phonon energy; $n^{\pm}_{{\bf
    q}\lambda}=n_{{\bf q}\lambda}+\frac{1}{2}\pm\frac{1}{2}$ with $n_{{\bf
    q}\lambda}$ representing the phonon number. For acoustic phonons, $\omega_{{\bf q}{\rm
    AC}}=v_{\rm ph}q$ with $v_{\rm ph}$ being the acoustic phonon velocity and
the scattering matrices are  
\begin{equation}
\hspace{-1cm}|M^{{\rm AC}\mu\mu^\prime}_{\bf
  k\nu,k+q\nu^\prime}|^2=\frac{D^2q}{2\rho_mv_{\rm ph}}I_{{\bf
    k}\nu,{\bf k+q\nu^\prime}}\delta_{\mu\mu^\prime},
\end{equation}
in which $D$ is the deformation potential and $\rho_m$ denotes the graphene mass
density.\cite{Chen,Hwang3} For the RI phonons,
\begin{equation}
|M^{{\rm RI}\mu\mu^\prime}_{\bf k\nu,k+q\nu^\prime}|^2=g
   \frac{v^2_F e^{-2qd}}{a q}\frac{I_{{\bf k}\nu,{\bf
         k+q\nu^\prime}}\delta_{\mu\mu^\prime}}{|1-V^0_{\bf
       q}\Pi({\bf q},\varepsilon_{{\bf k}\nu}-\varepsilon_{{\bf
         k+q}\nu^\prime})|^2},\label{MRI} 
\end{equation} 
where $g$ represents the dimensionless coupling parameter depending on the
material of the substrate,\cite{Perebeinos,Fratini}  $a$ is
the C-C bond distance, $d$ stands for the effective distance of the substrate
to the graphene sheet.\cite{Adam,Hwang2,Adam2,Fratini} 
For the optical phonons, 
\begin{equation}
\hspace{-1cm}|M^{{\rm OP}\mu\mu^\prime}_{\bf
  k\nu,k^\prime\nu^\prime}|^2=\frac{A^{\rm
    OP}_{\mu\mu^\prime\nu\nu^\prime{\bf k k^\prime}}}{2\rho_m\omega_{\rm OP}}.\label{MOP}
\end{equation}
In this investigation, we include the transverse optical phonons (${\rm K_{\rm
    TO}}$) near the
${\rm K}({\rm K}^\prime)$ point and the longitudinal (${\rm 
  \Gamma_{\rm LO}}$) as well as  transverse optical (${\rm \Gamma_{\rm TO}}$)
phonons near the ${\rm \Gamma}$ point. The corresponding parameters read
\begin{eqnarray}
&&\hspace{-0.8cm}A^{\rm \Gamma_{\rm
    LO}/\Gamma_{\rm TO}}_{\mu\mu^\prime\nu\nu^\prime{\bf k k^\prime}}\hspace{-0.05cm}=\hspace{-0.05cm}\langle
D^2_{\rm \Gamma}\rangle[1\hspace{-0.05cm}-\hspace{-0.05cm}\kappa\nu\nu^\prime\cos(\theta_{\bf k}+\theta_{\bf
  k^\prime}\hspace{-0.05cm}-\hspace{-0.05cm}2\theta_{\bf k^\prime-k})]\delta_{\mu\mu^\prime},\label{ALO}\\ 
&&\hspace{-0.8cm}A^{\rm K_{\rm
    TO}}_{\mu\mu^\prime\nu\nu^\prime{\bf k k^\prime}}\hspace{-0.05cm}=\hspace{-0.05cm}\langle
D^2_{\rm K}\rangle[1\hspace{-0.05cm}-\hspace{-0.05cm}\nu\nu^\prime\cos(\theta_{\bf k}-\theta_{\bf
  k^\prime})]\delta_{\mu,-\mu^\prime}\label{KTO}
\end{eqnarray}
with $\kappa=1(-1)$ for $\rm \Gamma_{\rm LO}$ ($\rm \Gamma_{\rm TO}$) phonons.\cite{Piscanec,Lazzeri}

In this paper, the dynamics of the RI and optical phonons are studied, while
the acoustic phonons are always set to be at the environment temperature $T_0$. We further
adopt the assumption following the previous 
work:\cite{Wang,Hale} the phonons in the same branch can equilibrate
themselves very quickly. Similar to those of the carriers, one can obtain the kinetic equations
of the hot phonons 
\begin{equation}
\partial_t{n}_{{\bf q}\lambda}=\left.\partial_t{n}_{{\bf q}\lambda}\right|_{\rm
  ep}+\left.\partial_t{n}_{{\bf q}\lambda}\right|_{\rm
  pp}.
\label{hotphonon}
\end{equation}
Here, $\left.\partial_t{n}_{{\bf q}\lambda}\right|_{\rm ep}$ and
$\left.\partial_t{n}_{{\bf q}\lambda}\right|_{\rm pp}$ come from the carrier-phonon
scattering and the anharmonic decay of hot phonons, respectively. For 
 RI and optical phonons, they are given by 
\begin{eqnarray}
&&\hspace{-1.cm}\left.\partial_t{n}_{{\bf q}\lambda}\right|_{\rm ep}=\frac{4\pi}{N_{\rm ph}}\sum_{{\bf
      kq^\prime}}\sum_{\nu\nu^\prime\mu\mu^\prime}\delta(\varepsilon_{{\bf
    k+q^\prime}\nu^\prime}-\varepsilon_{{\bf k}\nu}-\omega_{{{\bf q^\prime}}\lambda})\nonumber\\
&&\hspace{-0.cm}\mbox{}\times|M^{\lambda\mu\mu^\prime}_{\bf 
    k\nu,k+q^\prime\nu^\prime}|^2\big[f^>_{\mu{\bf k}\nu}f^<_{\mu^\prime{\bf
    k+q^\prime}\nu^\prime}(n_{{\bf q^\prime}\lambda}+1)\nonumber\\
&&\hspace{-0.cm}\mbox{}-f^<_{\mu{\bf k}\nu} f^>_{\mu^\prime{\bf k+q^\prime}\nu^\prime}
n_{{\bf q^\prime}\lambda}\big],\label{phononemissionrate}\\
&&\hspace{-1.cm}\left.\partial_t{n}_{{\bf q}\lambda}\right|_{\rm
  pp}=-\frac{n_{{\bf q}\lambda}-n^0_{{\bf q}\lambda}}{\tau_{pp}}. 
\end{eqnarray}
In above equations, $\tau_{pp}$ is the phenomenological relaxation time
from the phonon-phonon scattering; $n^0_{{\bf q}\lambda}$ is the number of the
$\lambda$ branch phonons at environment temperature $T_0$;
$N_{\rm ph}=m(E_{\rm max}/v_F)^2/4\pi$ is the number of the phonon
modes participating in the carrier-phonon scattering, 
where $E_{\rm max}$ represents the
upper energy of the hot carriers which are able to 
emit phonons\cite{Wang,Hale,interband}
 and $m=1$ for the RI, ${\rm \Gamma_{\rm TO}}$ and ${\rm \Gamma_{\rm LO}}$
phonons and $m=2$ for the two degenerate ${\rm K_{\rm TO}}$ phonons at the K and
K$^\prime$ point.

By numerically solving the kinetic equations [Eqs.~(\ref{KEE})
  and (\ref{hotphonon})] with the same numerical scheme laid out in
  Ref.~\onlinecite{Weng}, the temporal evolutions of the 
  carrier distribution and the phonon number can be obtained.  Then,
  the evolution of the optical transmission at the probe-photon energy 
$\omega_{\rm pr}$ can be calculated from $T_{\rm pr}(\omega_{\rm pr})=|1+N_{\rm 
  lay}\sigma(\omega_{\rm pr})\sqrt{\mu_0/\epsilon_0}/(1+n_{\rm ref})|^{-2}$
where $n_{\rm ref}$ is
the refractive index of the substrate and $N_{\rm lay}$ is the number of graphene
layers. The optical conductivity is given by $\sigma(\omega_{\rm
  pr})=-e^2(f_{\mu{\bf k_\omega}+}-f_{\mu{\bf
    k_\omega}-})/4$ with $|{\bf k_\omega}|=\omega_{\rm
  pr}/2v_F$.\cite{Dawlaty,Gusynin,Rana,Peres,Dawlaty2} The DT is then calculated
from $\Delta T_{\rm pr}/T^0_{\rm pr}=(T_{\rm pr}-T^0_{\rm pr})/T^0_{\rm pr}$, with
$T^0_{\rm pr}$ representing the transmission before pumping.  It is noted
that the epitaxial multilayer graphene can also be described by
our model, since they have been demonstrated to have similar
phononic and electronic properties to those of
single-layer graphene.\cite{Dawlaty,Latil,Varchon} 
Nevertheless, the differences between these two systems, i.e., the carrier-RI
phonon scattering becomes negligible when the number of layers is large, should
be taken into account. The material parameters used in our calculations
 are listed in Table~\ref{tab1}.

\begin {table}[tb]
 \caption{\label{tab1} Parameters used in the computation. 
}
 \begin{ruledtabular}
 \begin{tabular}{c|cc|cc}
   & $a$ & 1.42~\AA & $v_F$ & 1$\times10^8$~cm/s \\
   & $\omega_{\rm \Gamma}$ & 196~meV$^{\rm\,a}$ & $\langle D^2_{\rm
     \Gamma}\rangle$ & 45.60~eV$^2$/\AA$^2$$^{\rm\,a}$\\
   & $\omega_{K}$ & 161~meV$^{\rm\,a}$ & $\langle D^2_{K}\rangle$ &
   92.05~eV$^2$/\AA$^2$$^{\rm\,a}$ \\
   & $D$ & 19~eV$^{\rm\,b}$ & $v_{\rm ph}$ & 2$\times10^{6}$~cm/s$^{\rm\,b}$  \\
   & $\rho_m$ & 7.6$\times10^{-8}$~g/cm$^{2{\rm\,b}}$ & $d$ & 0.4~nm$^{\rm\,c,d}$\\
   \hline
 SiO$_2$  & $\omega_{\rm RI_1}$ & 59~meV$^{\rm\,d}$ & $g_1$ &
 $5.4\times10^{-3}$$^{\rm\,d}$\\
 & $\omega_{\rm RI_2}$ & 155~meV$^{\rm\,d}$ & $g_2$ & $3.5\times10^{-2}$$^{\rm\,d}$\\
 & $r_s$ & 0.8$^{\rm\,e}$ & $n_{\rm ref}$ & 1.5$^{\rm\,f}$   \\
   \hline
 SiC  & $\omega_{\rm RI}$  & 116~meV$^{\rm\,g}$ & $g$ & $1.4\times10^{-2}$$^{\rm\,g}$ \\
               & $r_s$ & 0.4$^{\rm\,d,i}$ & $n_{\rm ref}$ & 2.6$^{\rm\,h}$ \\
 \end{tabular}\\
 \end{ruledtabular}
\vskip 0.1cm
\begin{tabular}{lll}
$^{\rm a}$ Refs.~\onlinecite{Piscanec} and \onlinecite{Lazzeri}. &
$^{\rm b}$ Refs.~\onlinecite{Hwang3} and \onlinecite{Chen}. & $^{\rm
  c}$ Refs.~\onlinecite{Adam2}. \\ $^{\rm d}$
Ref.~\onlinecite{Fratini}. & $^{\rm e}$
Refs.~\onlinecite{Adam,Hwang}. & $^{\rm f}$
Ref.~\onlinecite{Ghosh}. \\ $^{\rm g}$
Ref.~\onlinecite{Perebeinos}. & $^{\rm i}$
Ref.~\onlinecite{Novikov}. & $^{\rm h}$ Ref.~\onlinecite{Levinshtein}
\end{tabular}
\end{table}

\section{RESULTS}
In this section we first study the
buildup  of the hot-carrier Fermi distribution in Sec.~IIIA.
We show that the Fermi distribution with identical temperature in the
conduction and valence bands can be established within 100~fs. 
Then, in Sec.~IIIB we simply use the hot-carrier Fermi distribution as
the initial carrier distribution and compare the calculated DT with the
experimental data. The evolutions of carrier and phonon temperatures 
  are also investigated here. Finally, we study the excitation-density and
  probe-photon--energy dependences of the slow DT relaxation rates in
  Sec.~IIIC.

\subsection{Buildup of hot-carrier Fermi distribution}

\begin{figure}[tbp]
  \includegraphics[width=7cm]{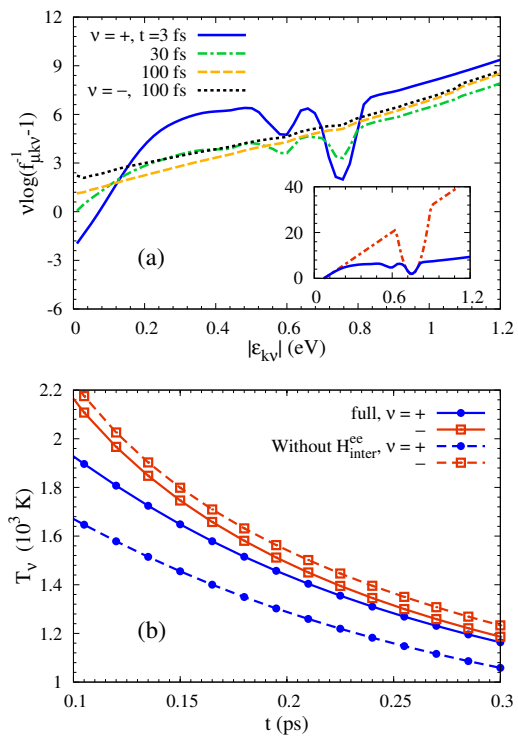}
  \caption{(Color online) (a) $\nu\log(f^{-1}_{\mu {\bf
      k}\nu}-1)$ as function of $|\varepsilon_{{\bf k}\nu}|$ for conduction-band
electrons ($\nu=+$) at $t=3$, 30 and 100~fs and valence band electrons
($\nu=-$) at $t=100$~fs. The results at $t=0$ and $3$~fs for conduction-band
electrons are plotted in the inset. (b) Temporal evolution of temperatures of
electrons in conduction (blue dots) and valence bands (red squares) from the
calculations with (solid curves) and without (dashed curves) the inter-band
Coulomb scattering $H^{\rm ee}_{\rm inter}$.}  
\label{figszw1}
\end{figure}

We set the initial carrier distribution to be
\begin{equation}
  f_{\mu{\bf k}\nu}=F(\varepsilon_{{\bf k}\nu})+\nu G(\varepsilon_{{\bf k}\nu}),
\end{equation}
in which $F(\varepsilon_{{\bf k}\nu})=1/\{1+\exp[(\varepsilon_{{\bf
    k}\nu}-\mu^0)/(k_BT_0)]\}$ is the carrier Fermi distribution before pumping
with $\mu^0$ denoting the initial chemical potential; 
$G(\varepsilon_{{\bf k}\nu})=A\exp\left[-{(|\varepsilon_{{\bf k}\nu}|-
    \omega_{\rm pu}/2)^2}/(2\Xi^2)\right]$ is the photo-generated
carrier distribution with $A$ and $\Xi$ representing the amplitude and standard
deviation, respectively and $\omega_{\rm pu}$ standing for the pump-photon energy.
In our computation here, the substrate is chosen to be SiO$_2$ and the equilibrium
carrier density $N_0$ is taken to be $6\times10^{11}$~cm$^{-2}$.
We also set $A=0.15$, $\Xi=19.3$~meV and $\omega_{\rm pu}=1.5$~eV,
corresponding to the excitation density $N_{ex}=8\times10^{11}$~cm$^{-2}$ 
and the absorbed intensity $I_a=1.9$~mJ/m$^2$. 
Then the conduction-band--electron
density is about $1.75$ times of the hole density (hole distribution is
defined as $f_{\mu {\bf k}h}\equiv 1-f_{\mu {\bf k}-}$).
The other parameters are taken to be $T_0=300$~K, 
$E_{\rm max}=0.9$~eV and $\tau_{pp}=3.8$~ps.
We first focus on the distribution of the
conduction-band electrons and plot the evolution of 
$\log(f^{-1}_{\mu{\bf k}+}-1)$ in Fig.~\ref{figszw1}(a). Note that if the
distribution $f_{\mu {\bf k}\nu}$ is the Fermi distribution, one has
\begin{equation}
\log(f^{-1}_{\mu {\bf k}\nu}-1)=(\nu v_{\rm F}k-\mu_\nu)/(k_BT_\nu),
\label{FitFermi}
\end{equation}
with $T_\nu$ and $\mu_\nu$ representing the temperature and chemical
potential in the corresponding band. Therefore, if the curves in the figure become linear
with $|\varepsilon_{{\bf k}\nu}|$, the buildup of the Fermi distribution is identified. 
In the inset of Fig.~\ref{figszw1}(a), one finds a valley located at
$|\varepsilon_{{\bf k}\nu}|=0.8$~meV in the initial distribution, coming from the
photo-generated carriers. With the evolution of time, the valley is rapidly smeared
out by the carrier-carrier Coulomb scattering as shown by the curves with $t=3$
and 30~fs in Fig.~\ref{figszw1}(a). Moreover, the curve with $t=100$~fs  becomes 
almost linear with $|\varepsilon_{{\bf k}\nu}|$, indicating the buildup
of the Fermi distribution. It is noted that this time scale is in the same
order as those in the experiments in graphite.\cite{Breusing} We stress that the
hot carrier Fermi distribution presented here is obtained directly from the 
microscopic kinetic equations and no ansatz is needed as all the
scatterings are included.

Then we turn to the valence-band electrons. Our calculations show that
the Fermi distribution of the valence-band electrons is
established in the same time scale as that of the conduction-band
ones, and we only plot 
$-\log(f^{-1}_{\mu{\bf k}-}-1)$ at $t=100$~fs in Fig.~\ref{figszw1}(a).
More importantly, one finds that the slopes for 
conduction (black dotted curve) and valence (yellow dashed curve) bands are very
close to each other at $t=100$~fs. This indicates that their corresponding 
temperatures $T_+$ and $T_-$ are very close to each other [see
Eq.~(\ref{FitFermi})], even though
the conduction-band--electron density is about $1.75$ times of the hole density.
To make this more pronounced, we plot the evolution of the hot-carrier temperatures $T_\nu$ fitted
from Eq.~(\ref{FitFermi}) in Fig.~\ref{figszw1}(b). The difference between
$T_+$ and $T_-$ is shown to be less than 10\%. This
phenomenon is due to the strong inter-band Coulomb scattering, which 
can be seen by comparing the temperatures from the calculations
with (solid curves) and without (dashed curves) the inter-band Coulomb
scattering $H^{\rm ee}_{\rm inter}$. 

\begin{figure*}[htb]
  \includegraphics[height=5.5cm]{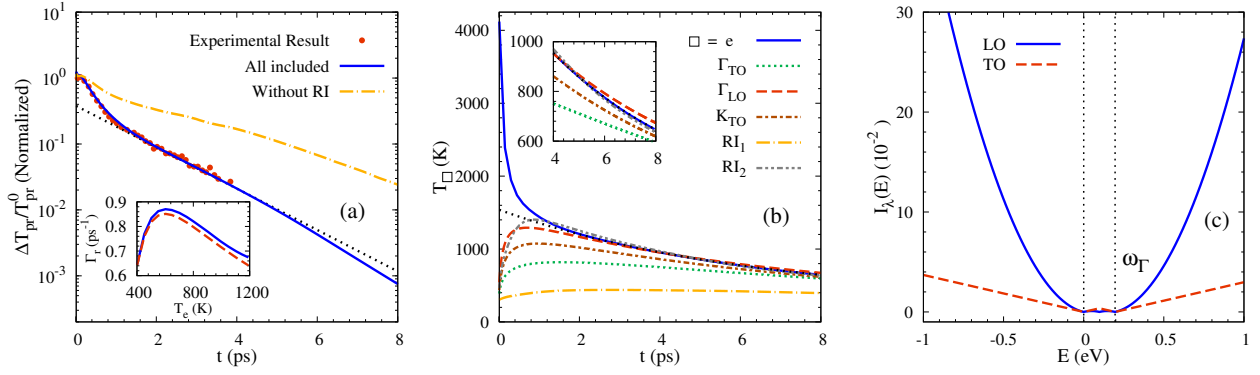}
\Black{\caption{(Color online) (a) DT from the numerical
  results compared with the experimental results in single-layer graphene on
  SiO$_2$ in Ref.~\onlinecite{Hale}. The calculation without the RI phonons
  (dash-dotted curve) is also plotted here. The
  results are normalized as Ref.~\onlinecite{Hale}. The black dotted curve represents the exponentially
  fitted curve of the DT in the time range of 2-4~ps. In the inset we plot 
  $\Gamma_r$ calculated from Eq.~(\ref{Gamma}) with (solid curve) and without 
  (dashed curve) the second term in the equation. (b) Temporal
  evolutions of the temperatures of carriers and phonons. Here the carrier temperature is shown to be fitted well
  by $T_0+T_A\exp(-\Gamma_Tt)$ (black dotted curve) for $t>1.5$~ps with
  $T_A=1240$~K and $\Gamma_T=0.16$~ps$^{-1}$. The inset zooms the time range
  4-8~ps. (c) Angular integration $I_{\lambda}(E)$ from Eq.~(\ref{Ieq}) for
 $\Gamma_{\rm LO}$ and $\Gamma_{\rm TO}$ phonons as function of the
  carrier energy $E$. The two black dotted lines indicate $E=0$ and
  $\omega_{\rm \Gamma}$, respectively.
}}
\label{figszw2}
\end{figure*}

\subsection{Temporal evolutions of DT and temperatures of carriers and phonons}

As shown in the previous subsection, the hot-carrier Fermi distribution is established very rapidly and the
temperatures of electrons in the conduction and valence bands are almost
identical. Therefore, in the following calculations, the 
initial carrier distribution is set to be 
\begin{equation}
f_{\mu{\bf k}\nu}(t=0)=1/\{1+\exp[(\varepsilon_{{\bf k}\nu}-\mu^0_\nu)/(k_BT^0_{e})]\},
\end{equation}
where $T^0_e$ denotes the hot-carrier temperature; $\mu^0_\nu$ represent the
chemical potentials in conduction ($\nu=+$) and valence ($\nu=-$) bands. $T^0_{e}$ and
$\mu^0_\nu$ can be determined by the equilibrium carrier density and the
excitation density $N_{ex}$ as well as the absorbed intensity $I_a$. For
simplicity, we set $I_a=N_{ex}\omega_{\rm pu}$ with $\omega_{\rm pu}$ denoting the
pump-photon energy. With this initial carrier distribution, the evolution of the
DT can be obtained numerically.

We first compare the DT from our calculations with the experimental results by Hale
{\it et al.}\cite{Hale} in single-layer graphene on SiO$_2$ substrates
[Fig.~\ref{figszw2}(a)].\cite{FitHale} Here $T_0=300$~K, $\omega_{\rm pr}=1.1$~eV and 
$\omega_{\rm pu}=1.5$~eV as indicated in the experiment. In this case, the excitation density is
much larger than the equilibrium carrier density, thus the chemical potential before the pumping has
little influence on the evolution of the DT and is set to be at the Dirac point for
simplicity. Then $T^0_e=4163$~K and $\mu^0_+=-\mu^0_-=-478$~meV correspond
  to $N_{ex}=4.6\times10^{12}$~cm$^{-2}$ and $I_a=11$~mJ/m$^{2}$, which are close to
the estimated values given in Ref.~\onlinecite{Hale}. The fitting parameters here
are $E_{\rm max}=0.9$~eV and $\tau_{pp}=3.8$~ps.  Our results agree very well with the
experimental data and show a fast relaxation with
 the characteristic time about 0.28~ps,
followed by a slow one with 1.33~ps.

To reveal the underlying physics of these
two relaxations, we plot the evolution of carrier and phonon
temperatures in Fig.~\ref{figszw2}(b). The carrier
temperature $T_e$ can be obtained by fitting $\log(f^{-1}_{\mu{\bf k}\nu}-1)$. The
temperature of hot phonons in $\lambda$ branch can be obtained from
$T_\lambda=\omega_{\lambda}/[k_B\ln(1+1/n_{{\bf q}\lambda})]$. From Fig.~\ref{figszw2}(b), it is seen that the
temperatures of the phonons first increase rapidly and then decrease slowly, with the peaks very
close to the crossover point between the fast and slow DT relaxations [see
Fig.~\ref{figszw2}(a)]. Since the fast increase of the phonon temperatures is due to
the rapid equilibration of the carrier-phonon system (less than $500$~fs)\cite{Breusing,Kampfrath,Butscher} and the 
decrease comes from the slow hot-phonon decay (about
several picosecond),\cite{Yan,Kang} the fast and slow relaxations of the DT can also be attributed to
these two processes, respectively.\cite{ElecAC}
This result supports the conjectures in the previous experimental
works.\cite{Hale,Wang,Hwang} In addition, by comparing the slow relaxation
of the DT with the exponential fitting curve [black dotted curve in
Fig.~\ref{figszw2}(a)], one finds that the relaxation rate slightly increases
with the temporal evolution when $t>4$~ps. This can be understood via the
approximate formula Eq.~(\ref{Gamma}),
which can be rewritten into
\begin{equation}
\Gamma_r=\Gamma_T\omega_{\rm pr}\frac{T_e-T_0}{2k_BT_e^2}-\frac{1}{k_BT_e}\frac{d\mu_+}{dt}
\end{equation}
by considering that $T_e$ can be fitted with 
  $T_e=T_0+T_A\exp^{-\Gamma_T t}$ for $t>1.5$~ps, as shown 
in Fig.~\ref{figszw2}(b).
We plot $\Gamma_r$ calculated from
this equation with and without the second
term in the inset of Fig.~\ref{figszw2}(a). It is seen that the
  first term in the equation is dominant and exhibits a peak at
  $T_e=2T_0=600$~K. In the time range
investigated here, $T_e$ is larger than 2$T_0$ and thus $\Gamma_r$ shows a
slight increase with increasing $t$ (decreasing
  temperature). Nevertheless, in the slow relaxation regime investigated 
  in the experiment, i.e., 2-4~ps, the exponential fit of DT is still 
  acceptable because the corresponding $\Gamma_r$ only changes by about 8\%.

Figure~\ref{figszw2}(b) also shows that the temperatures of phonons in
different branches are very different, originating from the different
carrier-phonon scattering strengths in different phonon branches. Interestingly,
the temperatures of the ${\rm \Gamma_{\rm TO}}$ and 
${\rm \Gamma_{\rm LO}}$ phonons differ much even though their carrier-phonon
scattering matrices are very similar [see Eq.~(\ref{ALO})].
This phenomenon comes from their different angular
dependences, which is neglected in the simple model in the experimental
works.\cite{Wang,Hale} To show this more clearly, we utilize the conditions that
the carrier distribution is isotropic and 
the distribution of optical phonons is
  independent of ${\bf q}$ (i.e., $n_{{\bf q}\lambda}=n_{\lambda}$), 
and rewrite Eq.~(\ref{phononemissionrate}) as 
\begin{eqnarray}
\left.\partial_t{n}_{\lambda}\right|_{\rm ep}&=&\frac{-4\pi}{N_{\rm ph}}\int^{\infty}_{-\infty}dE
I_\lambda(E)\big[f^<(E-\omega_\Gamma)f^>(E)n_{\lambda}\nonumber\\
&&\mbox{}-f^>(E-\omega_\Gamma)f^<(E)(n_{\lambda}+1)\big],
\end{eqnarray}
where $f^{>,<}(E)=\left.f^{>,<}_{\mu{\bf k}\nu}\right|_{\varepsilon_{\mu{\bf
      k}\nu}=E}$ and the angular integration $I_\lambda(E)$ is  given by
\begin{equation}
I_\lambda(E)=\hspace{-0.1cm}\sum_{{\bf k}\nu\mu\atop{\bf
    k^\prime}\nu^\prime\mu^\prime}\hspace{-0.1cm}|M^{{\lambda}\mu\mu^\prime}_{\bf
  k\nu,k^\prime\nu^\prime}|^2\delta(\varepsilon_{{\bf k^\prime}\nu^\prime}-E)\delta(\varepsilon_{{\bf k}\nu}-E+\omega_{\rm \Gamma}).
\label{Ieq}
\end{equation}
$I_{\lambda}(E)$ for ${\rm \Gamma_{\rm LO}}$ and ${\rm \Gamma_{\rm TO}}$
phonons are plotted as function of $E$ in 
Fig.~\ref{figszw2}(c). One finds that
$I_{\lambda}(E)$ for ${\rm \Gamma_{\rm LO}}$ (${\rm \Gamma_{\rm TO}}$) 
phonons is larger than the other one in the regime $E>\omega_{\rm \Gamma}$
and $E<0$ ($0<E<\omega_{\rm \Gamma}$). For the investigated excitation, $\omega_{\rm pu}$ is much
larger than $\omega_{\Gamma}$. Thus the intraband
carrier-phonon scattering (corresponding to $E>\omega_{\rm \Gamma}$ and $E<0$)
dominates the carrier-phonon thermalization. Consequently, the scattering
strength of ${\rm \Gamma_{\rm LO}}$ phonons is stronger  
and the corresponding temperature is higher. Another more interesting phenomenon
shown in Fig.~\ref{figszw2}(b) is that the carrier temperature can be even {\em lower} than
the hottest phonon one. This can be understood as follows: When the hot carriers are in
equilibrium with the hottest phonons, the cooling of the carrier is due to
the energy exchange with the other colder phonons; whereas the cooling of
the hottest phonons comes from the anharmonic decay of hot phonons. As shown above, the
carrier-phonon thermalization is faster than the hot-phonon decay. Thus the
temperature of carrier decreases faster and hence becomes lower than that of the
hottest phonons. We also discuss the contribution from the RI phonon, which
is neglected in the literature\cite{Wang,Hale} by plotting the
DT from the calculation without the RI phonons in Fig.~\ref{figszw2}(a).
It is seen that the exclusion of this phonon scattering makes a
marked difference, indicating that the RI phonons are very important to
the cooling process.

\begin{figure}[tp]
  \includegraphics[height=9cm]{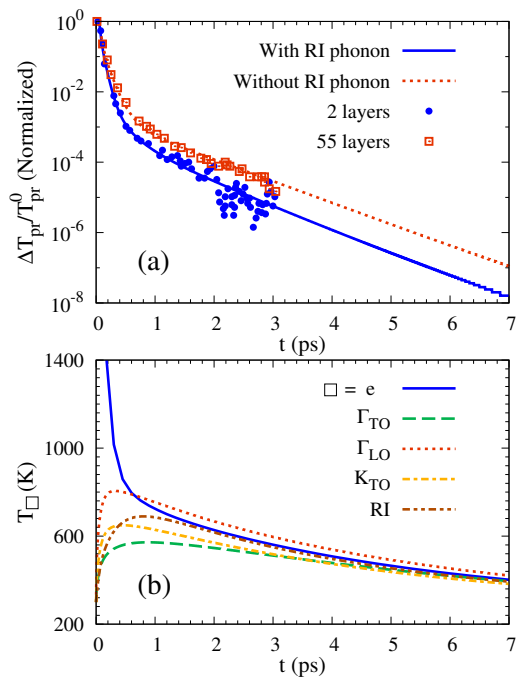}
\caption{(Color online) (a) Temporal evolution of the DT
  from the numerical results compared with the experimental data on SiC
  substrate extracted from Fig.~1 in Ref.~\onlinecite{Wang}. The results are
  normalized as that paper.
  (b) Temporal evolutions of the temperatures of
  carriers and phonons in the two-layer graphene. 
}
\label{figszw3}
\end{figure}
We then investigate the temporal evolutions of DT and temperatures of
  carriers and phonons in graphene on 6H-SiC. The results are compared with the
  experimental data reported by Wang {\it et al.}\cite{Wang} 
  (Fig.~1 in that paper) in Fig.~\ref{figszw3}(a).\cite{FitHale}
As mentioned above, the RI phonons are important 
only when the number of the graphene layers is small. Therefore, we include the
carrier--RI-phonon scattering for the two-layer sample (Sample C in
Ref.~\onlinecite{Wang}) and exclude it for the 55-layer
 sample (Sample A). As presented in Ref.~\onlinecite{Wang}, $\omega_{\rm pu}=\omega_{\rm
  pr}=1.6$~eV and the average equilibrium carrier densities for 55-layer and two-layer
 samples are taken to be 1$\times10^{11}$~cm$^{-2}$ and
6$\times10^{11}$~cm$^{-2}$, respectively. $N_{ex}=8.6\times10^{11}$~cm$^{-2}$ and
$I_a=2$~mJ/m$^2$ are consistent with the estimated values in
the experiment. Then the corresponding initial temperatures and chemical
  potentials are $T^0_e=4193$~K, $\mu^0_+=-1061$~meV and $\mu^0_-=1099$~meV for
  55-layer sample and $T^0_e=3512$~K, $\mu^0_+=-663$~meV and $\mu^0_-=825$~meV
  for two-layer sample. The fitting gives $E_{\rm max}=0.8$~eV and
  $\tau_{pp}=2.5$~ps.

Figure~3(a) shows good agreement between our calculations and the
experimental data in both samples, indicating that the
contribution from the RI phonons can be responsible for the difference
between the DTs in these two cases.  It is noted that to fit the results, the carrier--optical-phonon interaction
parameters $\langle D^2_{\rm \Gamma}\rangle$ and $\langle D^2_{\rm
  K}\rangle$ are chosen to be the same as
those adopted in Ref.~\onlinecite{Wang} but twice of those obtained from the
density functional calculations.\cite{Piscanec,Lazzeri} In fact, the values of
these parameters are still in
debate.\cite{Piscanec,Lazzeri,Rana2,Lazzeri2,Borysenko} Especially, the
influences of the electron-electron correlation\cite{Lazzeri2,Attaccalite} and the
interlayer coupling\cite{Borysenko2} on the carrier-phonon interaction
parameters in the epitaxial multilayer graphene are still unclear. Thus these
parameters can be sample dependent. We also show the evolution of carrier and
phonon temperatures in the two-layer graphene in Fig.~\ref{figszw3}(b). From our
results in Fig.~\ref{figszw3}(a) and (b), it can be seen that the
behaviors of carriers and phonons of graphene on a SiC substrate are similar
to those on a SiO$_2$ substrate, i.e., a fast DT relaxation of hundreds of
femtoseconds followed by a slower picosecond one as well as a lower carrier temperature
compared with the hottest phonons. In addition, the DT in the slow relaxation
regime decays exponentially in a large time range. This is because  $T_e$ is
around 2$T_0$ where $\Gamma_r$ varies mildly with $T_e$ as shown in the inset of
Fig~\ref{figszw2}(a).

\begin{figure}[tbp]
  \includegraphics[height=13.5cm]{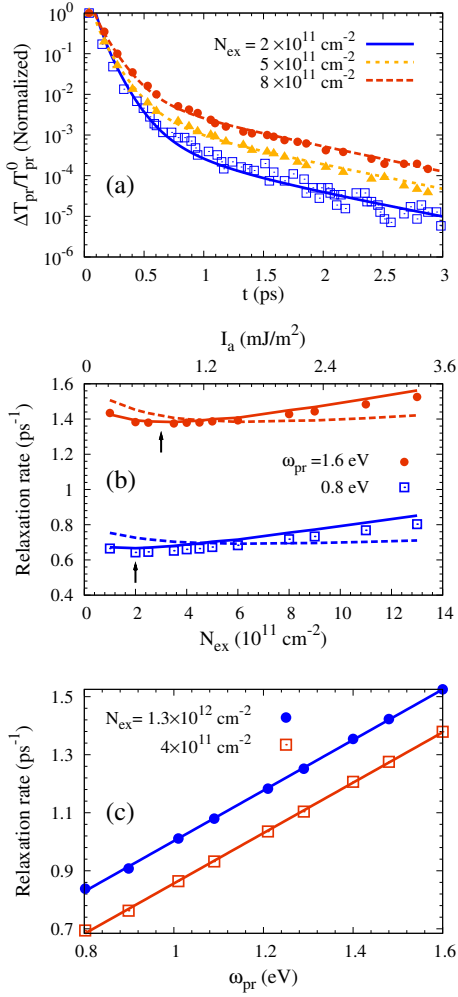}
  \caption{(Color online)  (a) DT from the numerical calculations compared with
    the experimental data [Fig.~3(b) in Ref.~\onlinecite{Wang}]. The pump pulse
    energies in the experiments are 2.5, 6 and 9.8~nJ from the bottom to the top. 
    (b) Slow relaxation rates of the DT  from the numerical calculations (dots) and the
    approximate formula Eq.~(\ref{Gamma}) (solid curves) as function of the
    excitation density $N_{ex}$ for different probe-photon energies. The DTs
    calculated from the first term in Eq.~(\ref{Gamma}) (dashed curves) are also plotted here. Note that the scale of
    $I_a$ corresponding to $N_{ex}$ is on top of the frame. The arrows show the
    positions of the valleys. (c) Slow relaxation rate of the DT as
    function of the probe-photon energy for different excitation densities. The
    results are fitted with linear functions (solid lines).
}   
\label{figszw4}
\end{figure}

\subsection{Excitation-density and probe-photon--energy
dependences of the slow  DT relaxation} 
In this subsection, we first compare the calculated DT  with the experimental data
extracted from Fig.~3(b) in Ref.~\onlinecite{Wang} for different excitation
densities in Fig.~\ref{figszw4}(a).\cite{FitHale}
Here the parameters are the same as those in Fig.~\ref{figszw3} unless otherwise
specified and the RI phonons are not included 
since the layer number of the sample is 16.
The fittings give the excitation densities $N_{ex}=2$, 5 and
8$\times10^{11}$~cm$^{-2}$ as well as $E_{\rm max}=0.6$, 0.7 and 0.8~eV
for the experimental results with pump-pulse energies being 
2.5, 6 and 9.8~nJ, respectively.
From this figure, one finds good agreement between our numerical results and
the experimental data for all three excitation densities.

By assuming that $E_{\rm max}$ is a linear function of $N_{ex}$ and fitting the
above values of $E_{\rm max}$ and $N_{ex}$, we obtain $E_{\rm max}= 0.033 
N_{ex} + 0.533$ ($E_{\rm max}$ and $N_{ex}$ are in units of eV and
$10^{11}$~cm$^{-2}$). With this relation, one can obtain the evolution of DT  
for other excitation densities. 
Here we concentrate on the relaxation rate in the time range of 2-3~ps, since the
evolution of DT in this regime shows a good exponential decay in the whole
excitation density range in this investigation. 
The results are plotted as dots in Fig.~\ref{figszw4}(b). It is seen that 
mild valleys appear in the $N_{ex}$ dependence and the excitation densities where
valleys appear tend to be lower for smaller $\omega_{\rm pr}$. 
To better understand this phenomenon, we also plot the results from the
approximate formula Eq.~(\ref{Gamma}) (solid curves) in
Fig.~\ref{figszw4}(b). Here $T_e$, $dT_e/dt$ and
$d\mu_+/dt$ in Eq.~(\ref{Gamma}) are chosen to be the ones at the middle of 
this time region, i.e., $t=2.5$~ps for each $N_{ex}$. One can see that the
results from Eq.~(\ref{Gamma}) agree very well with those from the kinetic
equations and exhibit valleys at the same $N_{ex}$.
The scenario of these valleys is as follows. 
With the increase of excitation density, the carrier and phonon temperatures
increase. Thus 
the electron-hole recombination and the cooling of carrier-phonon system
 both accelerate due to the enhanced carrier-phonon and
phonon-phonon scatterings. As a result,  $-d\mu_+/dt$ and
$-dT_e/dt$ in Eq.~(\ref{Gamma}) increase with $N_{ex}$.
Furthermore, our calculations show that $-d\mu_+/dt$ almost increases linearly
with $N_{ex}$ and $-dT_e/dt \sim N^{0.3}_{ex}$ in the excitation density range
investigated here. The excitation-density dependence of $T_e$ is more
complex: for $T_e\sim N^{\alpha}_{ex}$, $\alpha$ is around 0.18 for
$1\times10^{11}$~cm$^{-2}<N_{ex}<5\times10^{11}$~cm$^{-2}$ and then
decreases slowly with increasing $N_{ex}$ and reaches $0.1$ for
$N_{ex}=1.3\times10^{12}$~cm$^{-2}$. Therefore, 
the first term in Eq.~(\ref{Gamma}) first decreases and then increases with
increasing $N_{ex}$, while the second term increases monotonically with
$N_{ex}$. Under the joint effects of these two terms, the relaxation
rate shows a valley at the excitation density lower than that solely from the
first term [dashed curves in Fig.~\ref{figszw4}(b)].
Also by considering that the contribution of the first term decreases with
decreasing $\omega_{\rm pr}$, the valley moves to lower $N_{ex}$ 
when $\omega_{\rm pr}$ becomes smaller.
We also present more detail about the probe-photon--energy dependence in
Fig.~\ref{figszw4}(c). 
It is seen that the relaxation rate increases linearly with the increase of probe-photon 
energy $\omega_{\rm pr}$. This can also be understood via 
Eq.~(\ref{Gamma}) if one notices that $\mu_+$ and $T_e$ are independent of $\omega_{\rm pr}$.

\section{CONCLUSION}

In conclusion, we have microscopically investigated the dynamics
of nonequilibrium carriers and phonons in graphene by solving the kinetic equations with
the carrier-phonon and the carrier-carrier scatterings explicitly
included.  The hot-carrier Fermi distribution is found to be established within
100~fs. Furthermore, the temperatures of electrons in conduction and valence
bands are shown to be very close to each other even when the excitation 
density is comparable with the equilibrium carrier density. This is shown to be due to the
strong inter-band Coulomb scattering. 
Moreover, the temporal evolutions of the DT obtained from the kinetic equations
agree well with the experimental results\cite{Hale,Wang} for different
graphene layer numbers and excitation densities, with a fast relaxation about 
hundreds femtoseconds followed by a slow picosecond one presented. Based on
the results of the evolutions of carrier and phonon temperatures, we find that
the mechanisms leading to these two relaxations  
are the fast carrier-phonon thermalization and the hot-phonon
decay, respectively, which is consistent with the conjecture in the previous
experimental works.\cite{Wang,Hale,Huang} We also show that the
temperatures of the hot phonons in various branches are very 
different due to their different carrier-phonon scattering
strengths. Particularly, in spite of the similar carrier-phonon interaction
matrices, the scattering strengths of the TO and LO phonons
near the $\Gamma$ point are very different due to their different angular dependences.
In addition, the temperature of carriers can be lower than that of the hottest
phonons. This comes from the fact that the phonon temperatures are different for
different branches and the hot-phonon decay is unimportant during the phonon
thermalization. Our calculations also show that the contribution of the RI phonons
is important in the relaxation process.

Finally, we investigate the excitation-density and the probe-photon--energy dependences
of the slow DT relaxation rate. The relaxation rate is found to exhibit a mild valley in the
excitation density dependence. This phenomenon comes from the competition among
the increasing carrier temperature and 
the accelerating electron-hole recombination and carrier-phonon cooling
 with an increase of the excitation density. We also
show that the slow relaxation rate is linear with the probe-photon energy.

\begin{acknowledgments}
This work was supported by the National Natural Science Foundation of China
under Grant No.\ 10725417 and the National Basic Research Program of
China under Grant No.\ 2012CB922002. Two of the authors (B.Y.S. and Y.Z.) would like to
thank K. Shen for valuable discussions. 
\end{acknowledgments}

\appendix*

\section{Approximate formula of slow relaxation rate of DT}

We present the derivation of the approximate formula of the slow
relaxation rate of DT here.
In this time region, the Fermi distribution has been established. 
Thus the distribution functions
for electrons in the conduction and valence bands are
$f_c=\{\exp[\beta(\omega_{\rm pr}/2-\mu_+)]+1\}^{-1}$ and   
$f_v=\{\exp[\beta(-\omega_{\rm pr}/2-\mu_-)]+1\}^{-1}$, with
$\beta=1/(k_BT_e)$. Since $N_{\rm lay}\sigma(\omega_{\rm
  pr})\sqrt{\mu_0/\epsilon_0}/(1+n_{\rm ref})\ll 1$, 
\begin{eqnarray}
T_{\rm pr}&\approx&1+\frac{N_{\rm lay}e^2\sqrt{{\mu_0}/{\epsilon_0}}}{2(1+n_{\rm
    ref})}\left[\frac{1}{e^{\beta(\omega_{\rm
      pr}/2-\mu_+)}+1}\right.\nonumber\\
&&\mbox{}-\left.\frac{1}{e^{\beta(-\omega_{\rm pr}/2-\mu_-)}+1}\right].
\end{eqnarray}
Also considering that $\omega_{\rm pr}$ is much larger than $\mu_+$, $\mu_-$ and $k_BT_e$,
$\Delta T_{\rm pr}\equiv T_{\rm pr}-T^0_{\rm pr}$ can be expressed as 
\begin{equation}
\Delta T_{\rm pr}\approx\frac{e^2N_{\rm
    lay}\sqrt{\mu_0/\epsilon_0}}{2(1+n_{\rm ref})}e^{-\beta\omega_{\rm pr}/2}(e^{\beta\mu_+}+e^{-\beta\mu_-}).
\end{equation}
Thus, the relaxation rate of DT is given by
\begin{eqnarray}
 \hspace{-1cm}\Gamma_r&\equiv&-\frac{1}{\Delta T_{\rm pr}} \frac{d}{dt}(\Delta T_{\rm pr})\nonumber\\
 &\approx&\frac{\omega_{\rm pr}}{2}\frac{d\beta}{dt}-\beta\frac{d\mu_+}{dt}+\beta\frac{(\frac{d\mu_+}{dt}+\frac{d\mu_-}{dt})}{e^{\beta(\mu_++\mu_-)}+1}.
\end{eqnarray}
In the excitation density range investigated here, $|d\mu_+/dt+d\mu_-/dt|\ll
  |d\mu_+/dt|$. Therefore, one has 
\begin{equation}
\Gamma_r=-\frac{\omega_{\rm
    pr}}{2k_BT^2_e}\frac{dT_e}{dt}-\frac{1}{k_BT_e}\frac{d\mu_+}{dt}.
\label{Gamma}
\end{equation}


\begin{thebibliography}{0}
\bibitem{interband} Unlike Refs.~\onlinecite{Hale} and \onlinecite{Wang}, we do
not set the minimum value of $q$ participating in the carrier-phonon
scattering to be $\omega_{\lambda}/v_F$, since the phonons with 
$q<\omega/v_F$ can be heated by the inter-band carrier-phonon scattering.
\bibitem{FitHale} Since the pumping process is not simulated in this
  investigation, we shift the experimental results in
    Figs.~\ref{figszw2}(a), 3(a) and 4(a) by $60$, $30$ and $80$~fs, respectively.
\bibitem{ElecAC} Our calculations show that the electron-acoustic-phonon
  scattering is negligible for the evolution of the DT (not shown in the
  figure).
\end{thebibliography}
\end{document}